\documentclass[10pt,letterpaper]{article}
\usepackage{opex3}

\usepackage{graphicx}
\usepackage[intlimits]{amsmath}
\usepackage{amssymb}
\usepackage{amsthm}
\usepackage{booktabs}
\usepackage{multirow}
\usepackage{geometry}
\usepackage{bm}
\usepackage{slashbox}

\DeclareMathAlphabet{\mathcal}{OMS}{cmsy}{m}{n}

\begin{document}
 
\title{Surface scattering contribution to the plasmon width in embedded Ag nanospheres}
\author{R. Carmina Monreal,$^{1}$ S. Peter Apell,$^{2*}$ and Tomasz J. Antosiewicz$^{2,3}$ }
\address{$^{1}$Departamento de F\'{\i}sica Te\'orica de la Materia Condensada C5, and 
Condensed Matter Physics Center (IFIMAC),
Universidad Aut\'onoma de Madrid. E-28049 Madrid. Spain.\\
$^{2}$Department of Applied Physics and Gothenburg Physics Centre,\\ Chalmers University of Technology, SE-41296 G\"oteborg, Sweden \\
$^{3}$Centre of New Technologies, University of Warsaw, Banacha 2c, 02-097 Warsaw, Poland
}
\email{$^*$apell@chalmers.se}

\begin{abstract}
Nanometer-sized metal particles exhibit broadening of the localized surface plasmon resonance (LSPR) in comparison to its value predicted by the classical Mie theory. Using our model for the LSPR dependence on non-local surface screening and size quantization, we quantitatively relate the observed plasmon width to the nanoparticle radius $R$ and the permittivity of the surrounding medium $\epsilon_m$. For Ag nanospheres larger than 8 nm only the non-local dynamical effects occurring at the surface are important and, up to a diameter of 25 nm, dominate over the bulk scattering mechanism. Qualitatively, the LSPR width is inversely proportional to the particle size and has a nonmonotonic dependence on the permittivity of the host medium, exhibiting for Ag a maximum at $\epsilon_m\approx2.5$. Our calculated LSPR width is compared with recent experimental data.
\end{abstract}


%\bibliographystyle{osajnl}
%\bibliography{../../../Nasze_publikacje/TJA_bibliography_2014.03.14}

\section{Introduction}
Since the early 1980s calculations and measurements on the optical properties of ultra-small particles have seen  important progress \cite{PRL_52_1925_ekardt, PRB_33_2828_borensztein}. Coming from the angstrom end of the size spectrum we see an equally profound development in the field of clusters and the region where the two meet is especially interesting [3-5].%\cite{PRA_48_R1749_liebsch, PRB_58_6748_yannouleas, PRB_47_10706_fredrigo}.
Underlying this development are of course both the advance in computational tools \cite{NL_12_429_townsend} as well as increasingly refined experimental techniques. Especially  methods which make it possible to produce and study individual particles of a well-defined shape and size underpin  the rapid development of the field [7-10]. %\cite{PRL_88_077402_sonnichsen, NJP_4_93_sonnichsen, PRL_80_4249_klar, PNAS_107_14530_peng}.
At the same time the possible applications of plasmonic resonances in these systems, ranging from medical therapies, via biosensing  to new devices are but a few examples of what is driving the field rapidly forward [11-14].%\cite{Sci_337_1072_ciraci, PRL_108_106802_fernandez, PRL_110_263901_teperik, OE_20_25201_apell}.

Despite the positive development described above there are still crucial aspects of plasmonic resonances in small particles that are not completely understood and hence not utilized to their full potential. 
In a recent paper \cite{NJP_15_083044_carmina} we presented a theoretical model for analyzing the size dependence of the surface plasmon resonance of metallic nanospheres in a range of sizes down to a few nanometers. Within this model, we explicitly showed how different microscopic mechanisms affect the energy of the surface plasmon in quantitative agreement with recent published experimental results for Ag and Au.

In this article we address a  question related to the one considered in \cite{NJP_15_083044_carmina}. We investigate the behavior of the width of the surface plasmon resonance as a function of the size of the particle and the influence of its surrounding medium, because the understanding of the plasmon decay rate is essential to control its spectral response.
Whereas the energy shift is small compared to the peak position in an absorption spectrum of a small particle, the width is a much more sensitive parameter to what takes place inside and around the particle. Mechanisms affecting the width include shape changes, vibrations in the particle (with electron-phonon coupling dependent on particle geometry \cite{OER_22_36_fan}) and scattering processes of the conduction electrons common to bulk materials (electron-electron and electron-defect scattering). 
Chemical effects associated with the bonding of adsorbed molecules from a reactive host matrix to the particle surface \cite{SurfSci_281_153_persson} or the presence of ligand molecules also contribute to the plasmon width. Radiation damping is not significant for the very small particles we consider in our paper. However, of major concern for us is another damping mechanism, the so called surface scattering related to the non-harmonic part of the surface potential where for a perfect harmonic confining potential there is no associated damping but now the direct excitation of electron-hole pairs in an electronically very inhomogeneous region of space (known as spill-out region) plays a major role as we will show below. 
A basic question to address in this context is to what extent is the collective mode broadening an intrinsic or extrinsic property of the small particle. This is an essential issue because it limits the electromagnetic field enhancement at its surface.

We calculate the optical  extinction cross-section of Ag spheres of nanometric size embedded in different host media, characterized by their permittivity $\epsilon_m$, assumed to be frequency independent. We take a top-down approach to the nanoparticles and, starting from a large size, make them smaller and smaller. For metal spheres less than 25 nm in diameter, the wavelengths of interest are typically much larger than the size so that a quasi-static dipolar approximation is valid to calculate the optical properties.
Within this approximation, the surface scattering mechanism for the damping of the surface plasmons is introduced in the theory using the framework presented in \cite{NJP_15_083044_carmina}, which includes non-local surface screening and size quantization effects, and was used to reproduce the LSPR shift in frequency. The surface scattering mechanism gives a surface plasmon linewidth which depends on size and also on the permittivity of the surrounding medium. In Ag, this width reaches a maximum as a function of the host permittivity, independently of size.
We also analyze the quality factor of the surface plasmon resonance of these Ag spheres. We find that, when surface scattering is the dominant mechanism for the damping of the surface plasmon, the quality factor shows a trend with size different than when radiative damping dominates. Our study is motivated by recent quantitative experimental determination of the size dependence of the surface plasmon resonance in single Ag nanoparticles coated with SiO$_2$ \cite{NL_9_3463_baida}.
In previous experiments \cite{TCA_116_514_cottancin, PRB_57_1963_palpant}, the samples contained ensembles of particles with a dispersion in size which made  a detailed investigation of this issue difficult. The remarkable agreement we find between the experimental and our calculated plasmon widths affirms our approach.

\section{Theory} 
For the small particles we are interested in, the extinction cross-section can be written as a function of frequency $\omega$ as
\begin{equation}
\sigma(\omega)=4 \pi \frac{\omega}{c} \sqrt{\epsilon_m} \mathrm{{\,Im[}} \alpha(\omega){]},
\label{sigma}
\end{equation}
where $c$ is the speed of light, $\alpha$ is the particle polarizability, and the surrounding medium is non-magnetic. In the classical Mie theory \cite{AnnPhys_25_377_mie}, the polarizability of a small metal sphere of radius $R$ (diameter $D$) and dielectric function $\epsilon(\omega)$ is dominated by the electric dipole term which is given by
\begin{equation}
\alpha_M(\omega)=R^3 \frac{\epsilon(\omega)-\epsilon_m}
{\epsilon(\omega)+2\epsilon_m}.
\label{Mie-polarizability}
\end{equation}

Separating $\epsilon(\omega)$ into its real and imaginary components, $\epsilon(\omega)=\epsilon_1(\omega)+i \epsilon_2(\omega)$, 
the imaginary part of Eq.~(\ref{Mie-polarizability}) is
\begin{equation}
\mathrm{{Im[}} \alpha_M(\omega){]}=3 \;R^3 \; \epsilon_m \frac{\epsilon_2(\omega)}
{(\epsilon_1(\omega)+2\epsilon_m)^2+ (\epsilon_2(\omega))^2},
\end{equation}
which gives to Eq.~(\ref{sigma}) a nearly Lorentzian line-shape, with a peak at the frequency $\omega_M$ (the classical local surface plasmon frequency) a full width at half maximum $\Gamma_M$ fulfilling, 
\begin{align}
\epsilon_1(\omega_M)+2\epsilon_m&=0 &\mathrm{and} & &\Gamma_M=&2 \frac{\epsilon_2(\omega_M)}{\left[\frac{\partial \epsilon_1(\omega)}{\partial \omega}\right]_{\omega_M}},
\refstepcounter{equation}
\tag{\theequation a,b}
\end{align}
respectively.

The dielectric function of noble metals is usually written as the sum of the contributions of the d electrons $\epsilon_{d}(\omega)$ and the sp electrons, the second one being of the Drude form, as
\begin{equation}
\epsilon(\omega)=\epsilon_{d}(\omega)- \frac{\omega_{p}^{2}}{\omega^2+i \omega/\tau_b},
\label{epsilon-w}
\end{equation}
where $\omega_{p}$ is the free-electron plasma frequency and $\tau_b$ is the bulk relaxation time. Then, the position and full width at half-maximum of the local surface plasmon resonance are obtained from Eqs.~(4a) and (4b) as ($\omega_M \tau_b \gg 1$)
\begin{align}
\omega_M&=\frac{\omega_{p}}{\sqrt{\mathrm{{Re}}[\epsilon_d(\omega_M)]+2 \epsilon_m}} & \mathrm{and} & & \Gamma_M=& \frac{\frac{1}{\tau_b} +\mathrm{{Im}}[\epsilon_d(\omega_M)] \frac{\omega_{M}^{3}}{\omega_{p}^{2}} }
{1+\frac{1}{2}\frac{\omega_{M}^{3}}{\omega_{p}^{2}}\left[\frac{\partial \mathrm{{Re}}[\epsilon_d(\omega)]}{\partial \omega}\right]_{\omega_M}},
%\label{w-Mie}
\refstepcounter{equation}
\tag{\theequation a,b}
\end{align}
respectively. Eqs.~(6a) and (6b) show that this classical theory produces resonances which are independent of particle size in the quasistatic approximation. However, it is well known experimentally that the frequency and width of the surface plasmons are size-dependent for particles of radii which, in principle, are within the quasistatic regime. In particular, the width increases with decreasing size linearly in $\frac{1}{R}$. 
This fact has been phenomenologically taken into account in the classical formulation by assuming that a relaxation time $\tau$ should be used in Eq.~(6b) which has contributions from the usual inelastic scattering of the electrons in the bulk, $\tau_b$, and also from the scattering at the boundaries, according to the expression $\frac{1}{\tau}= \frac{1}{\tau_b}+ g_s \frac{v_{F}}{R}$, where $v_{F}$ is the Fermi velocity and $g_s$ is usually taken as an adjustable parameter \cite{PRB_48_18178_vollmer}. 
A surface contribution to the relaxation time of the form expressed by the second term on the right hand side of the preceeding equation has been computed by several authors using the so-called quantum-box model, in which the electronic states are quantized in an infinite barrier potential at the surface of the particle [23-26].% \cite{JPSJ_21_1765_kubo, SSComm_18_385_ascarelli, PRB_11_2871_ruppin, Plas_9_185_mortensen}.

In this paper we introduce the surface scattering mechanism for plasmon damping  using the theory presented in \cite{NJP_15_083044_carmina}. In this framework, non-local, dynamical effects occurring at the surface of the metal particle,  where electrons spill-out, are incorporated in the theory by means of the complex length $d_r(\omega)$, defined as \cite{PS_26_113_apell}
\begin{equation}
\frac{d_r(\omega)}{R}=\frac{ \int dr\; r(R-r)\delta\rho(r,\omega)}
{ \int dr\; r^2 \delta\rho(r,\omega)},
\label{dr-general}
\end{equation}
where $\delta\rho(r, \omega)$ is the induced electronic density, $r$ is the radial coordinate and the integrals extend to the whole space. The real part of $d_r(\omega)$ is related to the position of the center of gravity of the screening charge density at the surface and produces red/blue shifts of the surface plasmon frequency of spheres depending on whether the screening charge sits outside/inside the jellium edge.
The imaginary part of $d_r(\omega)$ describes the absorption of energy from an external probe creating surface excitations (surface plasmons and surface electron-hole pairs) and  the sign criterium used to define Eq.~(\ref{dr-general}) implies that $\mathrm{{Im}} [-d_r(\omega)]$ is positive. The fact that $\mathrm{{Im}} [-d_r(\omega)]$ is finite at the frequency of the surface plasmon indicates that this collective excitation has a finite lifetime because it is coupled and decays into the incoherent excitation of electron-hole pairs (Landau damping). These surface effects can be incorporated into the polarizability of small spheres, yielding the generalization of the classical Eq.~(\ref{Mie-polarizability}) as \cite{PS_26_113_apell}
\begin{equation}
\alpha(\omega)=R^3 \frac{\left(\epsilon(\omega)-\epsilon_m\right)\left(1-\frac{d_r(\omega)}{R}\right)}
{\epsilon(\omega)+2\epsilon_m+2\left(\epsilon(\omega)-\epsilon_m\right) \frac{d_r(\omega)}{R}}.
\label{polarizability}
\end{equation}

In the classical Mie theory, where the charge density is  proportional to $\delta(r-R)$, $d_r(\omega)$ is zero and these surface effects are absent. In our approach Eq.~(\ref{epsilon-w}) is generalized in order to include quantum size effects as
\begin{equation}
\epsilon(\omega)=\epsilon_{d}(\omega)- \frac{\omega_{p}^{2}}{\omega^2-\Delta^2+i \omega/\tau_b},
\label{epsilon}
\end{equation}
with  $\Delta$ playing the role of an energy gap introduced by the quantization of the electronic levels of the nanoparticle. We should point out that the results we are going to present below are very little influenced by the effects of quantization due to size. This is because, as discussed in \cite{NJP_15_083044_carmina}, quantum size effects are only important for spheres smaller than about 5 nm in diameter, while the experimental widths we will compare to are measured in spheres of diameter on the order of and larger than 10 nm. 

The surface contribution to the absorption cross section can be analyzed in a simple way by assuming that $\epsilon(\omega)$ is real, which is a  good approximation for Ag in the range of frequencies of interest below the onset for interband transitions (less than 4 eV) where $\mathrm{{Re}}[\epsilon_{d}(\omega)] \gg \mathrm{{Im}}[\epsilon_d(\omega)]$
and $\omega \tau_b \gg 1$.
Then the imaginary part of Eq.~(\ref{polarizability}) reads
\begin{equation}
\mathrm{Im}[\alpha(\omega)]=   
\frac {3 R^3 \epsilon(\omega) \left(\epsilon(\omega)-\epsilon_m\right)\frac{-\mathrm{Im}[d_r(\omega)]}{R}}
{\left[\epsilon(\omega)+2\epsilon_m+2\left(\epsilon(\omega)-\epsilon_m\right) \frac{\mathrm{Re}[d_r(\omega)]}{R}\right]^2 
+\left[2\left(\epsilon(\omega)-\epsilon_m\right)\frac{\mathrm{Im}[d_r(\omega)]}{R}\right]^2}.
\label{Im-alpha}
\end{equation}

Therefore, for large values of $R$, Eq.~(\ref{Im-alpha}) yields an absorption cross section which has a nearly Lorentzian shape, having a maximum at the surface plasmon frequency $\omega_s$, and is given by
\begin{equation}
\epsilon(\omega_s)+2\epsilon_m+2\left(\epsilon(\omega_s)-\epsilon_m\right) \frac{\mathrm{{Re}}[d_r(\omega_s)]}{R}=0,
\label{wsp}
\end{equation}
and full width at half-maximum $\Gamma_s$ 
\begin{equation}
\Gamma_s \simeq 4\frac{\left(\epsilon(\omega_s)-\epsilon_m\right)} 
{\left[\frac{\partial \epsilon_1(\omega)}{\partial \omega}\right]_{\omega_s}}\frac{\mathrm{Im}[d_r(\omega_s)]}{R}
\simeq 12 \frac{\epsilon_m}
{\left[\frac{\partial \epsilon_1(\omega)}{\partial \omega}\right]_{\omega_s}}\frac{\mathrm{Im}[-d_r(\omega_s)]}{R},
\label{gammasp}
\end{equation}
where $\epsilon_1(\omega)$ denotes the real part of Eq.~(\ref{epsilon}). Hence, this model gives a width of the surface plasmon resonance which increases linearly with $\frac{1}{R}$ with a slope proportional to $\mathrm{{Im}}[-d_r(\omega_s)]$. A more detailed analysis, using for $\epsilon(\omega)$ the complex form of Eq.~(\ref{epsilon}), shows that these surface effects will dominate over the classical bulk contribution to the width when $2 \frac{\mathrm{{Im}}[-d_r(\omega_s)]}{R} \geq \frac{1}{\omega_s \tau_b}$. Using typical values, $\hbar\omega_s \simeq 3$~eV, $ \mathrm{{Im}}[-d_r(\omega_s)]  \simeq 0.2$~nm and  $\frac{\hbar}{\tau_b}\simeq 0.02$ eV, this relation is well satisfied for Ag spheres smaller than 30~nm in diameter, the sizes of interest in this work.

It is interesting for the analysis of the results that we will present below to obtain the limit of Eq.~(\ref{gammasp}) for very large values of the permittivity of the surrounding medium. Assume  $\epsilon_m$ is so large that the surface plasmon frequency is in the region where $\epsilon_d(\omega_s) \ll  2 \epsilon_m$ and  $\omega_s \simeq \omega_p/ \sqrt{2\epsilon_m}$. Then
\begin{equation}
\left[\frac{\partial \epsilon_1(\omega)}{\partial \omega}\right]_{\omega_s} \simeq 2 \frac{\omega_{p}^2}{\omega_{s}^3} \simeq \frac{2 (2\epsilon_m)^{\frac{3}{2}}}{\omega_{p}},
\end{equation} 
and, in the limit $\epsilon_m \gg 1$, eq. (\ref{gammasp}) reads
\begin{equation}
\lim_{\epsilon_m \rightarrow \infty} \Gamma_s = 3 \frac{\omega_p}{\sqrt{2\epsilon_m}} \frac{\mathrm{{Im}}[-d_r(\omega_s)]}{R}.
\label{gamma-infty}
\end{equation}

Moreover, since the phase space for exciting electron-hole pairs decreases as $\omega$ for $\omega\ll\omega_p$, one has
\begin{equation}  
\lim_{\omega \rightarrow 0} \mathrm{{Im}}[-d_r(\omega)] \longrightarrow 0. 
\label{dr-infty}
\end{equation}
Actually, it is an exact result for a planar surface that $ \mathrm{{Im}}[-d_{\perp}(\omega)] \simeq \omega$ in the low frequency limit $\omega\ll\omega_p$ 
 \cite{PRB_27_6058_persson, PRB_36_7378_liebsch}.
Hence, from Eqs.~(\ref{gamma-infty}) and (\ref{dr-infty}), the surface scattering (Landau damping) contribution to the linewidth of the surface plasmon has to vanish for large values of $\epsilon_m$.

Following \cite{NJP_15_083044_carmina} we substitute $d_r(\omega)$ by its corresponding counterpart for a planar surface $d_{\perp}(\omega)$
because it has been shown \cite{PRL_52_1925_ekardt} that the induced charge density at the surface of a sphere is very similar to that of a planar surface down to a few nanometers in size. This complex quantity has been calculated for surfaces of free-electron like metals in contact with vacuum, using either  the Time Dependent Local Density Approximation (TDLDA) \cite{PRB_36_7378_liebsch} or the Random Phase Approximation (RPA) \cite{ProgSurfSci_12_287_feibelman} to the non-local self-consistent dielectric response  of the metal electrons confined by a Lang-Kohn potential barrier.  We adapt these calculations to our case of Ag by means of the renormalization of the bulk plasma frequency to $\omega_{p}^{*}$ given by $\mathrm{{Re}} [\epsilon(\omega_{p}^{*})]=0$ \cite{ PRB_33_2828_borensztein,  NJP_15_083044_carmina}. In this way we introduce the effects of d-electrons in $d_r$. 
This approximation is motivated by three physical facts: (i) the 4d-electrons of Ag are localized and largely excluded from the surface region where the conduction electrons spill-out. This is an essential consideration for obtaining the observed blue shift of the surface plasmons in planar Ag surfaces and Ag spheres \cite{PRB_48_11317_liebsch}.  (ii) When plotted versus the reduced frequency $\omega/\omega_{p}$,  
the function $d_{\perp}(\omega/\omega_{p})$ is qualitatively similar for all simple metals, from Al ($r_s \simeq 2 a_0$) to Cs ($r_s\simeq 6 a_0$) \cite{PRB_36_7378_liebsch, ProgSurfSci_12_287_feibelman}.
(iii) In noble metals,  $\omega_p^*$ (instead of the free-electron $\omega_p$) is the frequency at which the metal changes behavior from surface screening to penetration of an external electromagnetic field.

\section{Results}

\begin{figure}
\centering
\includegraphics{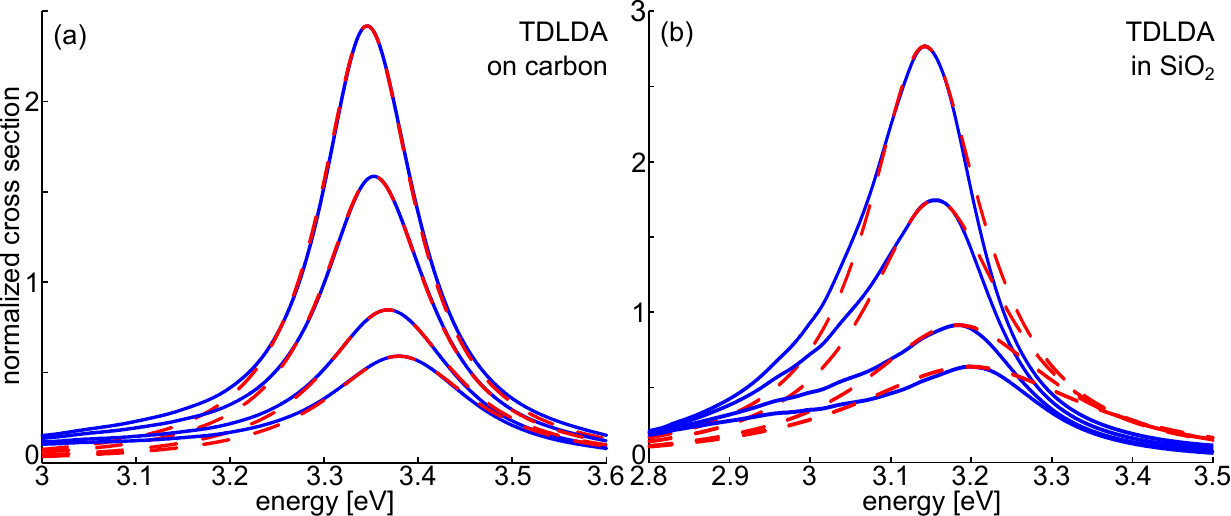}
\caption{Absorption cross-sections (continuous blue lines) of silver spheres  embedded in: (a) a host medium of dielectric function 1.5 (average value for spheres on a carbon substrate) and (b) a host medium of dielectric function 2.16 (SiO$_2$). Calculations have been performed using the Time-Dependent Local Density Approximation for  the dielectric response of the electronic density of Ag in the surface region. The cross-sections are normalized to the surface area of the sphere, the radius being  10 nm, 7.5 nm, 5 nm and 4 nm from top to bottom in each subfigure. Both show a blue-shift of the plasmon for decreasing size and a broadening of the peak. Dashed red lines are Lorentzians constructed from the width and peak position of the full calculation. They are a very good match to the spectra in (a) and not so good in (b). This reflects the detailed structure of the surface response function which is the major contribution to the absorption cross-section for
  these small spheres.}
\label{Figure_absorptance}
\end{figure}

The absorption cross-section is calculated from Eqs. (\ref{sigma}), (\ref{polarizability}) and (\ref{epsilon}), using the TDLDA and RPA calculations of $d_{\perp}(\omega)$ for $r_s=3 a_0$  (appropriate for Ag) from Refs. \cite{PRB_36_7378_liebsch} (Figure 2 therein) and \cite {ProgSurfSci_12_287_feibelman} (Figures 9 and 10 therein), respectively.  The RPA values of  $\mathrm{Im}[-d_{\perp}(\omega)]$ are also plotted in Figure 2(b) of Ref. \cite{PRB_36_7378_liebsch}. Here, $r_s$ is the one-electron radius and $a_0$ denotes the Bohr radius. Both approximations can account for the surface plasmon relation of dispersion at planar surfaces of the free-electron-like metals, even tough  TDLDA gives an overall better agreement with experiment \cite{SS_247_302_tsuei}. We find that the results we present and discuss in this section do not depend on the self-consistent treatment of the dielectric response as long as the electronic spill-out of the metal electrons is include
 d in the theory. 
We will see that they are able to account for the experimental magnitude of the surface plasmon width of Ag nanospheres in SiO$_2$, thus reinforcing the importance of an appropriate treatment of the surface electronic density. $\epsilon(\omega)$ and $\frac{1}{\tau_b}$ are taken from the experimental data of Johnson and Christy \cite{PRB_6_4370_johnson} from where $\frac{\hbar}{\tau_b}=0.016$ eV. The width of the surface plasmon resonance,  $\Gamma_R$, is defined as the full width at half maximum of the corresponding absorption curve.

Figure~\ref{Figure_absorptance}  shows absorption cross-sections of Ag spheres of decreasing sizes, calculated  using the TDLDA values of $d_{\perp}(\omega)$, embedded in (a) a host medium of $\epsilon_m=1.5$ and (b) SiO$_2$ ($\epsilon_m=2.16$). For the sake of clarity, the cross-sections are normalized to the surface area $4 \pi R^2$. Also plotted are Lorentzians (dashed lines) having the same peak position and width as the calculated cross-sections, and normalized to the same height. The value $\epsilon_m=1.5$ was chosen in \cite{NJP_15_083044_carmina} for analyzing the blue shift in frequency of the surface plasmon  in Ag spheres on a carbon substrate.
Notice in Fig.~\ref{Figure_absorptance}(a) the blue shift of the surface plasmon peak and the width of the resonance increasing with decreasing size. The Lorentzian fit is in excellent agreement in these cases. For the case of spheres embedded in SiO$_2$ shown in Fig.~\ref{Figure_absorptance}(b), the resonances show a smaller blue shift and are much wider than in Fig.~\ref{Figure_absorptance}(a). 
%As a reference, a sphere of $R=4$~nm in carbon has a surface plasmon resonance shifted in energy by 0.05 eV with a linewidth of 0.22 eV while the same sphere in SiO$_2$ shows a surface plasmon with half the energy shift and is  50$\%$ wider.
As the permittivity of the host medium increases, the frequency of the surface plasmon decreases moving to a region of frequencies where $\mathrm{Re}[d_{\perp}(\omega)]$ is smaller (less blue shift) and $\mathrm{Im}[-d_{\perp}(\omega)]$ is larger (more width). One should also note that the absorption cross section is less symmetric around the maximum and therefore not so well fitted by a Lorenztian (red dashed lines). This is due to the fact that $\mathrm{Im}[d_{\perp}]$ is a strongly varying function of frequency in this range of values of $\frac{\omega}{\omega_{p}^{*}}$.
In general, we find that the absorption cross-section becomes more asymmetric with decreasing size in both RPA (not shown) and TDLDA calculations, as the resonance gets wider and, consequently, the functional dependence of $\mathrm{Im}[d_{\perp}(\omega)]$ becomes more important. Therefore, the energy dependence of the surface excitations is a source of  skewness for the optical absorption cross-section of perfectly spherical particles with  radii typically smaller than 5 nm.
Even though the dispersion in frequencies of $\epsilon(\omega)$ is another source of asymmetry for the cross-section, we find that surface scattering is more important for Ag, were the resonance is below the onset of interband transitions. This would be different in the case of Au.

\begin{figure}
\centering
\includegraphics{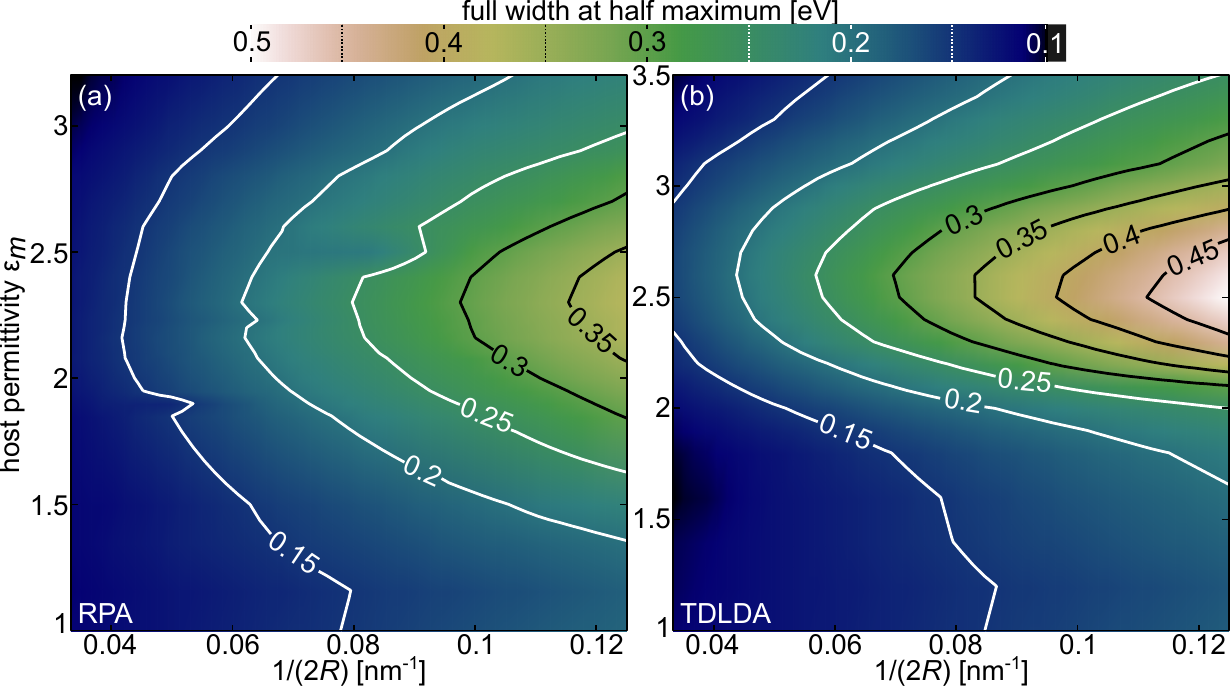}
\caption{Color plot of the surface plasmon width as a function of inverse size and host permittivity for (a) the Random Phase Approximation  and (b) the Time-dependent Local Density Approximation. The contours and the scale on top are in eV. Results are dominated by the surface effect we calculate. In both cases the width scales directly with the inverse size. Notice that the width is maximum for a host dielectric medium with permittivity 2.4-2.6.}
\label{Figure_colormap}
\end{figure}

Figure~\ref{Figure_colormap} displays color maps of the surface plasmon width as function of the permittivity of the host medium and inverse of size, using (a) RPA and (b) TDLDA values of $d_{\perp}(\omega)$. Except at the largest sizes and lowest permittivities, where the width is on the order of 0.1 eV, the values shown in these figures are completely dominated by surface effects (Landau damping). For a given $\epsilon_m$, the width of the surface plasmons increases almost linearly with $R^{-1}$  with a slope which increases with $\epsilon_m$ up to a maximum, as can be appreciated in this figure.
This behavior is completely different from the classical theory, where the numerator of Eq.~(6b) decreases as $\epsilon_m$ increases, yielding decreasing values of $\Gamma_M$ in the range of values of $\epsilon_m$ shown in these figures. We should mention here that the experimental results for Ag clusters in different surroundings compiled in Fig.~3 of Ref.~\cite{PRB_48_18178_vollmer} seem to confirm the trend of the present calculations.
Then, for all values of  $R$, our model predicts that the plasmon width has a maximum as a function of the host permittivity at $\epsilon_m \simeq 2.4- 2.6 $, depending on the approximation to $d_{\perp}(\omega)$, RPA or TDLDA. The existence of this maximum can be understood from our analysis above showing that the surface scattering mechanism for plasmon damping tends to vanish for sufficiently large values of $\epsilon_m$.

\begin{figure}
\centering
\includegraphics{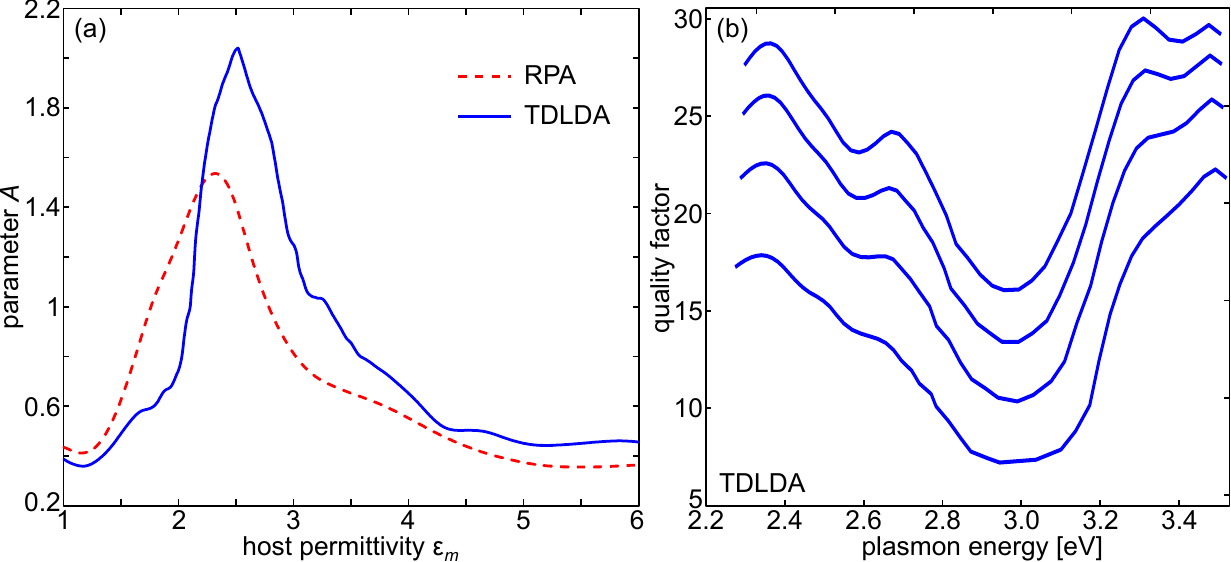}
\caption{(a) The linear coefficient of the plasmon width in $v_F/R$, parameter $A$, as a function of the host permittivity for the Random Phase Approximation (dashed red line) and the Time-Dependent Local Density Approximation (solid blue line). The maximum of A leads to the maximum in the surface plasmon damping mechanism present in both approximations.
(b) Quality factor of the surface plasmon in silver spheres for  the TDLDA. Radius being 10 nm, 7.5 nm, 5 nm and 4 nm from top to bottom. We take the quality factor as resonance energy divided by width. To obtain the resonance energies, we changed the host dielectric permittivity in the range 1 - 6. Again we see how the maximum plasmon width comes in around 3  eV giving a minimum in the quality factor.}
\label{Figure_AQ}
\end{figure}

Our findings can be seen more clearly when we fit our calculated width to the usual form 
\begin{equation}
\Gamma_{R}=\Gamma_0+A \frac{v_F}{R},
\end{equation}
where $\Gamma_0$ and $A$ are the fitting parameters. The values of $\Gamma_0$ we obtain are nearly the same for both theories and very close to the values of $\Gamma_M$ given by Eq. (6b). These values are in the range  $0.04 - 0.09$ eV and, in general, are small in comparison with the full width as can be appreciated in Fig.~\ref{Figure_colormap}. Figure~\ref{Figure_AQ}(a) shows the parameter $A$ as a function of $\epsilon_m$ for the RPA and TDLDA type of calculations. It shows a pronounced maximum, since $A$ is roughly proportional to $\mathrm{Im}[-d_{\perp}(\omega_s)]$, which yields the maximum of plasmon width with host permittivity seen in Fig.~\ref{Figure_colormap}.
We note here that $A$ is larger than 1 for certain values of host dielectric function in sharp contrast to what is found by box models ignoring the spill-out effect of the surface electronic density. It is worth noticing that the values of $A$ extracted from both theories can differ by almost a factor of 2 at some values of $\epsilon_m$ and therefore an experimental determination of $A$ could be used  as a test for these theories.

Figure~\ref{Figure_AQ}(b) displays the quality factor $Q$ for Ag spheres of decreasing sizes, calculated using the TDLDA  values of $d_{\perp}(\omega)$. The  quality factor gives a measure of the local field enhancement at the particle surface when the plasmon mode is excited \cite{PRL_88_077402_sonnichsen} and is defined as $Q=\omega_R/ \Gamma_R$, where $\omega_R$ and $\Gamma_R$ are the peak position and full width at half maximum of the plasmonic resonances. The  curves are obtained by continuously changing the host permittivity in the range $1 \le \epsilon_m \le 6$, obtaining the resonance energies shown in the figure.
The important feature in this figure is the minimum in $Q$ at $\omega_R \simeq 3$ eV, coincident with the maximum of the plasmon width discussed above, for all sizes considered in this work. This is completely different from the behavior of $Q$ when radiative damping dominates ($R>15$ nm), where $Q$ decreases  quickly with $\omega_R$ \cite{JQSRT_114_45_kolwas}. Another important difference seen in $Q$ in the regimes of surface scattering versus radiative damping is its behavior with particle size. When surface scattering dominates ($R<10$ nm) $Q$ decreases with $R$ because $\Gamma_R$ is nearly proportional to $R^{-1}$ while $\omega_R$ depends weakly on $R$. 
However, in the radiative damping region, $Q$ decreases with increasing size \cite{JQSRT_114_45_kolwas}. The small features in $Q$ are also present in a calculation neglecting the surface scattering mechanism and are brought about by the experimental values of $\epsilon(\omega)$. We should also mention that a calculation of $Q$ excluding the effects of surface scattering will produce quality factors bigger than the ones shown in this figure by factors of 3-8. The same trends are obtained in calculations using the RPA to the dielectric response. This fact points out to the importance of using realistic surface models when calculating optical properties of small metallic systems. 

\begin{figure}
\centering
\includegraphics{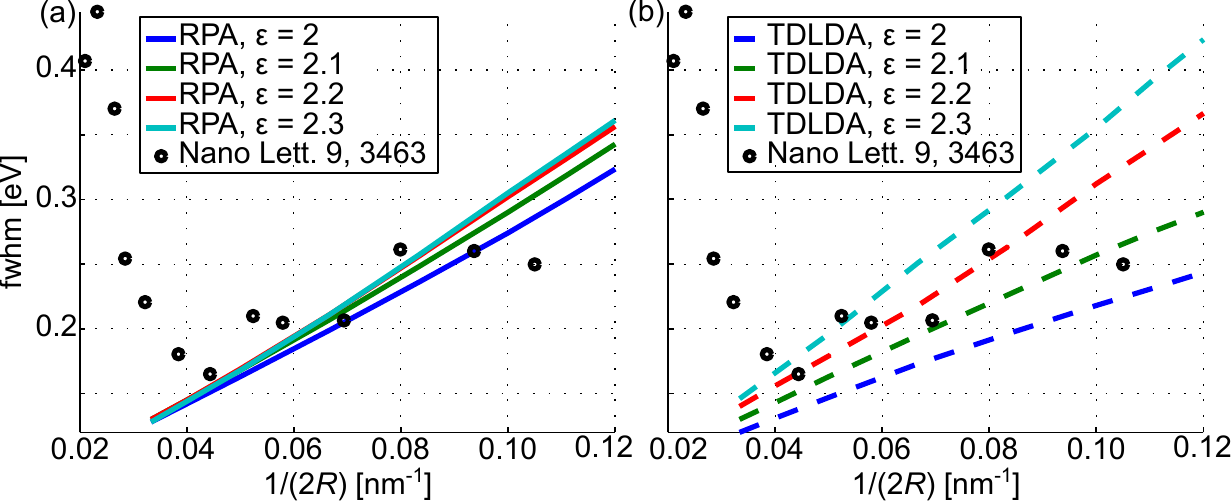} %[width=12.5cm]
\caption{The surface plasmon width as a function of inverse size for (a) the Random Phase approximation  and (b) the Time-dependent Local Density Approximation. Results are given for four different permittivities of the host medium differing from the value for SiO$_2$ by less than 10$\%$, and compared to the experimental results for Ag spheres coated with that material reported in \cite{NL_9_3463_baida} (circles). Notice the radiative damping taking over the physics as we approach particles larger than 25 nm. Then, the smallest damping is obtained for sizes of ca. 20 nm.}
\label{Figure_comparison_to_experiment}
\end{figure}

Figure~\ref{Figure_comparison_to_experiment} shows our calculated widths, using (a) RPA  and (b) TDLDA  values of $d_{\perp}(\omega)$, compared with the experiments in \cite{NL_9_3463_baida} for single Ag nanoparticles coated with a 15 nm shell of SiO$_2$. We show results for values of the host permittivity differing by less than 10$\%$ from the nominal value for this material, $\epsilon_m=2.16$. This is because it is probable that the thickness of the coating is not enough to assume an infinite host in all the cases or that the coating might not be uniformly distributed around the metal core.
The sharp increase in the experimental width for $2R>25$ nm has been shown in \cite{NL_9_3463_baida, JQSRT_114_45_kolwas} to be due to radiative damping, which becomes important for increasing sizes, as is clearly illustrated in Fig.~8 of Ref.~\cite{JQSRT_114_45_kolwas}. The agreement we find with the experiment, at the sizes where the quasi-static dipolar approximation is valid, is noticeable. Even though the calculated widths have a stronger variation with $\epsilon_m$ when using the TDLDA values of $d_{\perp}(\omega)$  [Fig. \ref{Figure_comparison_to_experiment}(b)] than when using the RPA values [Fig. \ref{Figure_comparison_to_experiment}(a)], one can appreciate that this experiment can be reproduced by both approximations, using reasonable values of $\epsilon_m$ and without further adjustable parameters. 

Recently Lerm\'e et al. \cite{JPCL_1_2922_lerme, JPCC_115_14098_lerme} have analyzed the effects of confinement and  permittivity of the host matrix on the width of the surface plasmons. These authors take a bottom-up approach to the nanoparticles and perform full TDLDA calculations of the optical absorption cross-section of Ag clusters of diameters smaller that 11 nm, in vacuum ($\epsilon_m=1$) and embedded in SiO$_2$ ($\epsilon_m=2.16$) and Al$_2$O$_3$ ($\epsilon_m=3.2$). 
Therefore in these works the surface plasmon damping mechanism is Landau damping, as in our case. Due to quantization of the electronic states in the cluster, the absorption spectrum is fragmented into multiple peaks whose widths can be changed by means of a line broadening parameter. It is found that, for the larger clusters considered, the linewidth increases in a way approximately linear with $\frac{1}{R}$, with a slope that increases with the permittivity of the host medium, from vacuum to Al$_2$O$_3$. This is the same trend we find in our TDLDA calculations. 
%  even though for the modeling of the clusters considered in \cite{JPCL_1_2922_lerme, JPCC_115_14098_lerme} the maximum of the linewidth would happen at $\epsilon_m >3.2$.
When the results for Ag in SiO$_2$ in \cite{JPCL_1_2922_lerme, JPCC_115_14098_lerme} are extrapolated linearly up to the region of the experimental sizes, the magnitude of the width so obtained is much smaller than ours and in poorer agreement with the experiment. Nevertheless, when decreasing the size below 5 nm, we know that quantum size effects are essential to account for the color of the energy shift the surface plasmons and, consequently, they would also affect the surface plasmon width. This is the region of the size spectra where first principles calculations are indeed necessary.

\section{Conclusions}
In this work we have theoretically analyzed the evolution with size and host permittivity of the linewidth of the local surface plasmon resonances in spherical Ag nanoparticles, using the same framework which allowed us to successfully reproduce the shift in energy of these dipolar modes. The physical magnitude controlling that evolution for both energy position and width of the surface plasmons is the complex ratio $d_r(\omega_s)/R$, whose real and imaginary parts  basically fix the energy and width of the surface plasmons respectively.
We approximate $d_r(\omega)$ by its counterpart for a planar surface, calculated using two different approximations (RPA and TDLDA) to the non-local self-consistent dielectric response  of the metal electrons confined by a Lang-Kohn potential barrier. We find that the damping of the surface plasmon collective mode in Ag spheres with diameters smaller than ca. 25 nm,  is dominated by the surface scattering mechanism that couples this mode to single electron-hole pairs excitations.
The linewidth of the plasmons in these spheres increases with decreasing size in a way approximately linear in $1/R$, with a slope which depends on the permittivity of the host medium, showing a maximum at $\epsilon_m \simeq 2.4-2.6$, depending on the approximation to $d_r(\omega)$. This fact strongly affects the quality factor of the plasmonic resonances which is minimum at resonance energies of ca. 3~eV, independent of particle size.
From our calculations, we extract the linear coefficient of the plasmon width in $v_F/R$, which depends on the surrounding permittivity, and  could be used to account for this increased damping in a phenomenological manner. We compare our results with a recent experimental determination of the surface plasmon resonance damping of single Ag spheres coated with SiO$_2$, finding a good agreement in the range of sizes $ 2R < 25$ nm where the dipolar approximation is valid. We thus conclude that our approach to surface plasmons in small particles is a reliable one down to 10 nm in diameter, a range of sizes where sophisticated first principles calculations are not feasible yet.
However, these kinds of calculations are necessary when decreasing the particle size below 5 nm, where quantum size effects play an important role in accounting for the color of the energy shift of the surface plasmons and should be equally important for a determination of their widths. Moreover, their extension to larger sizes would be highly interesting.

\section*{Acknowledgments}
RCM acknowledges financial support from the Spanish Mineco via the project FIS2011-26516. TJA thanks the Foundation of Polish Science for support via the project HOMING PLUS/2013-7/1. TJA and SPA acknowledge support from the Swedish Foundation for Strategic Research via the Functional Electromagnetic Metamaterials for Optical Sensing project SSF~RMA~11.


\begin{thebibliography}{10}
\newcommand{\enquote}[1]{``#1''}

\bibitem{PRL_52_1925_ekardt}
W.~Ekardt, \enquote{Dynamical polarizability of small metal particles:
  Self-consistent spherical jellium background model,} Phys. Rev. Lett.
  \textbf{52}, 1925--1928 (1984).

\bibitem{PRB_33_2828_borensztein}
Y.~Borensztein, P.~De~Andr\`es, R.~Monreal, T.~Lopez-Rios, and F.~Flores,
  \enquote{Blue shift of the dipolar plasma resonance in small silver particles
  on an alumina surface,} Phys. Rev. B \textbf{33}, 2828--2830 (1986).

\bibitem{PRA_48_R1749_liebsch}
J.~Tiggesb\"aumker, L.~K\"oller, K.-H. Meiwes-Broer, and A.~Liebsch,
  \enquote{Blue shift of the mie plasma frequency in {A}g clusters and
  particles,} Phys. Rev. A \textbf{48}, R1749--R1752 (1993).

\bibitem{PRB_58_6748_yannouleas}
C.~Yannouleas, \enquote{Microscopic description of the surface dipole plasmon
  in large {N}a$_{N}$ clusters $(950\lesssim {N} \lesssim 12050)$,} Phys. Rev.
  B \textbf{58}, 6748--6751 (1998).

\bibitem{PRB_47_10706_fredrigo}
S.~Fedrigo, W.~Harbich, and J.~Buttet, \enquote{Collective dipole oscillations
  in small silver clusters embedded in rare-gas matrices,} Phys. Rev. B
  \textbf{47}, 10706--10715 (1993).

\bibitem{NL_12_429_townsend}
E.~Townsend and G.~W. Bryant, \enquote{Plasmonic properties of metallic
  nanoparticles: The effects of size quantization,} Nano Letters \textbf{12},
  429--434 (2012).

\bibitem{PRL_88_077402_sonnichsen}
C.~S\"onnichsen, T.~Franzl, T.~Wilk, G.~von Plessen, J.~Feldmann, O.~Wilson,
  and P.~Mulvaney, \enquote{Drastic reduction of plasmon damping in gold
  nanorods,} Phys. Rev. Lett. \textbf{88}, 077402 (2002).

\bibitem{NJP_4_93_sonnichsen}
C.~S\"onnichsen, T.~Franzl, T.~Wilk, G.~von Plessen, and J.~Feldmann,
  \enquote{Plasmon resonances in large noble-metal clusters,} New J. Phys.
  \textbf{4}, 93 (2002).

\bibitem{PRL_80_4249_klar}
T.~Klar, M.~Perner, S.~Grosse, G.~von Plessen, W.~Spirkl, and J.~Feldmann,
  \enquote{Surface-plasmon resonances in single metallic nanoparticles,} Phys.
  Rev. Lett. \textbf{80}, 4249--4252 (1998).

\bibitem{PNAS_107_14530_peng}
S.~Peng, J.~M. McMahon, G.~C. Schatz, S.~K. Gray, and Y.~Sun,
  \enquote{Reversing the size-dependence of surface plasmon resonances,} Proc.
  Natl. Acad. Sci. U.S.A. \textbf{107}, 14530--14534 (2010).

\bibitem{Sci_337_1072_ciraci}
C.~Cirac\`i, R.~T. Hill, J.~J. Mock, Y.~Urzhumov, A.~I.
  Fern\'andez-Dom\'inguez, S.~A. Maier, J.~B. Pendry, A.~Chilkoti, and D.~R.
  Smith, \enquote{Probing the ultimate limits of plasmonic enhancement,}
  Science \textbf{337}, 1072--1074 (2012).

\bibitem{PRL_108_106802_fernandez}
A.~I. Fern\'andez-Dom\'inguez, A.~Wiener, F.~J. Garc\'ia-Vidal, S.~A. Maier,
  and J.~B. Pendry, \enquote{Transformation-optics description of nonlocal
  effects in plasmonic nanostructures,} Phys. Rev. Lett. \textbf{108}, 106802
  (2012).

\bibitem{PRL_110_263901_teperik}
T.~V. Teperik, P.~Nordlander, J.~Aizpurua, and A.~G. Borisov, \enquote{Robust
  subnanometric plasmon ruler by rescaling of the nonlocal optical response,}
  Phys. Rev. Lett. \textbf{110}, 263901 (2013).

\bibitem{OE_20_25201_apell}
A.~Garcia-Etxarri, P.~Apell, M.~K\"all, and J.~Aizpurua, \enquote{A combination
  of concave/convex surfaces for field-enhancement optimization: the indented
  nanocone,} Opt. Express \textbf{20}, 25201--25212 (2012).

\bibitem{NJP_15_083044_carmina}
R.~C. Monreal, T.~J. Antosiewicz, and S.~P. Apell, \enquote{Competition between
  surface screening and size quantization for surface plasmons in
  nanoparticles,} New J. Phys \textbf{15}, 083044 (2013).

\bibitem{OER_22_36_fan}
Y.~Fan, J.~Li, H.~Chen, X.~Lu, and X.~Liu, \enquote{Size-dependence of the
  effective electron-phonon energy relaxation in hollow gold nanospheres,}
  Opto-Electron. Rev. \textbf{22}, 36--40 (2014).

\bibitem{SurfSci_281_153_persson}
B.~N.~J. Persson, \enquote{Polarizability of small spherical metal particles:
  influence of the matix environment,} Surf. Sci. \textbf{281}, 153--162
  (1993).

\bibitem{NL_9_3463_baida}
H.~Baida, P.~Billaud, S.~Marhaba, D.~Christofilos, E.~Cottancin, A.~Crut,
  J.~Lerm\'e\/, P.~Maioli, M.~Pellarin, N.~Del~Fatti, F.~Vall\'ee,
  A.~S\'anchez-Iglesias, I.~Pastoriza-Santos, and L.~M. Liz-Marz\'an,
  \enquote{Quantitative determination of the size dependence of surface plasmon
  resonance damping in single {A}g@{S}i{O}$_{2}$ nanoparticles,} Nano Lett.
  \textbf{9}, 3463--3469 (2009).

\bibitem{TCA_116_514_cottancin}
E.~Cottancin, G.~Celep, J.~Lerm\'e, M.~Pellarin, J.~Huntzinger, J.~Vialle, and
  M.~Broyer, \enquote{Optical properties of noble metal clusters as a function
  of the size: Comparison between experiments and a semi-quantal theory,}
  Theoretical Chemistry Accounts \textbf{116}, 514--523 (2006).

\bibitem{PRB_57_1963_palpant}
B.~Palpant, B.~Pr\'evel, J.~Lerm\'e, E.~Cottancin, M.~Pellarin, M.~Treilleux,
  A.~Perez, J.~L. Vialle, and M.~Broyer, \enquote{Optical properties of gold
  clusters in the size range 2--4 nm,} Phys. Rev. B \textbf{57}, 1963--1970
  (1998).

\bibitem{AnnPhys_25_377_mie}
G.~Mie, \enquote{Beitr\"age zur {O}ptik tr\"uber {M}edien, speziell kolloidaler
  {M}etall\"osungen,} Ann. Phys. \textbf{25}, 377--445 (1908).

\bibitem{PRB_48_18178_vollmer}
H.~H\"ovel, S.~Fritz, A.~Hilger, U.~Kreibig, and M.~Vollmer, \enquote{Width of
  cluster plasmon resonances: Bulk dielectric functions and chemical interface
  damping,} Phys. Rev. B \textbf{48}, 18178--18188 (1993).

\bibitem{JPSJ_21_1765_kubo}
A.~Kawabata and R.~Kubo, \enquote{Electronic properties of fine metallic
  particles. ii. plasma resonance absorption,} Journal of the Physical Society
  of Japan \textbf{21}, 1765--1772 (1966).

\bibitem{SSComm_18_385_ascarelli}
P.~Ascarelli and M.~Cini, \enquote{Red shift of the surface plasmon resonance
  absorption by fine metal particles,} Solid State Commun. \textbf{18},
  385--388 (1976).

\bibitem{PRB_11_2871_ruppin}
R.~Ruppin, \enquote{Optical properties of small metal spheres,} Phys. Rev. B
  \textbf{11}, 2871--2876 (1975).

\bibitem{Plas_9_185_mortensen}
A.~V. Uskov, I.~E. Protsenko, N.~A. Mortensen, and E.~P. O'Reilly,
  \enquote{Broadening of plasmonic resonance due to electron collisions with
  nanoparticle boundary: a quantum mechanical consideration,} Plasmonics
  \textbf{9}, 185--192 (2014).

\bibitem{PS_26_113_apell}
P.~Apell and A.~Ljungbert, \enquote{A general non-local theory for the
  electromagnetic response of a small metal particle,} Physica Scripta
  \textbf{26}, 113--118 (1982).

%\bibitem{PRL_30_975_feibelman}
%P.~J. Feibelman, \enquote{Sensitivity of surface-plasmon dispersion and damping
%  to potential barrier shape,} Phys. Rev. Lett. \textbf{30}, 975--978 (1973).

\bibitem{PRB_27_6058_persson}
B.~N.~J. Persson and P.~Apell, \enquote{Sum rules for surface response functions with application to 
the van der Waals interaction between an atom and a metal,} 
Phys. Rev. B. \textbf{27}, 6058--6065 (1983).

\bibitem{PRB_36_7378_liebsch}
A.~Liebsch, \enquote{Dynamical screening at simple-metal surfaces,} Phys. Rev.
  B \textbf{36}, 7378--7388 (1987).

\bibitem{ProgSurfSci_12_287_feibelman}
P.~J. Feibelman, \enquote{Surface electromagnetic fields,} Prog. Surf. Sci.
  \textbf{12}, 287--407 (1982).
  
  \bibitem{PRB_48_11317_liebsch}
  A.~Liebsch, \enquote{Surface-plasmon dispersion and size dependence of Mie resonance: Silver versus simple metals,} Phys. Rev.
  B \textbf{48}, 11317--11328 (1993).


\bibitem{SS_247_302_tsuei}
K.-D. Tsuei, E.~W. Plummer, A.~Liebsch, E.~Pehlke, K.~Kempa, and P.~Bakshi,
  \enquote{The normal modes at the surface of simple metals,} Surf. Sci.
  \textbf{247}, 302--326 (1991).

\bibitem{PRB_6_4370_johnson}
P.~Johnson and R.~Christy, \enquote{Optical constants of the noble metals,}
  Phys. Rev. B \textbf{6}, 4370--4379 (1972).

\bibitem{JQSRT_114_45_kolwas}
K.~Kolwas and A.~Derkachova, \enquote{Damping rates of surface plasmons for
  particles of size from nano- to micrometers; reduction of the nonradiative
  decay,} J. Quant. Spectrocs. Radiat. Transfer \textbf{114}, 45--55 (2013).

\bibitem{JPCL_1_2922_lerme}
J.~Lerm\'e, H.~Baida, C.~Bonnet, M.~Broyer, E.~Cottancin, A.~Crut, P.~Maioli,
  N.~Del~Fatti, F.~Vall\'ee, and M.~Pellarin, \enquote{Size dependence of the
  surface plasmon resonance damping in metal nanospheres,} J. Phys. Chem. Lett.
  \textbf{1}, 2922--2928 (2010).

\bibitem{JPCC_115_14098_lerme}
J.~Lerm\'e, \enquote{Size evolution of the surface plasmon resonance damping in
  silver nanoparticles: Confinement and dielectric effects,} J. Phys. Chem. C
  \textbf{115}, 14098--14110 (2011).

\end{thebibliography}
\end{document}